\documentclass[12pt,dvips]{article}
\usepackage{epsfig}
\begin{document}

\title{\bf Inhomogeneous structure formation may alleviate need for accelerating universe}

\date{}

\author{Johan Hansson \& Jesper Lindkvist \\ Department of Physics \\ Lule{\aa} University of Technology \\ SE-971
87 Lule\aa, Sweden}

\maketitle

\begin{abstract}
When taking the real, inhomogeneous and anisotropic matter
distribution in the semi-local universe into account, there may be
no need to postulate an accelerating expansion of the universe
despite recent type Ia supernova data. Local curvatures must be
integrated (over all space) to obtain the global curvature of the
universe, which seems to be very close to zero from cosmic
microwave background data. As gravitational structure formation
creates bound regions of positive curvature, the regions in
between become negatively curved in order to comply with a
vanishing global curvature. The actual dynamics of the universe is
altered due to the self-induced inhomogeneities, again more
prominently so as structure formation progresses. Furthermore,
this negative curvature will increase as a function of time as
structure formation proceeds, which mimics the effect of ``dark
energy" with negative pressure. Hence, the ``acceleration" may be
merely a mirage. We make a qualitative and semi-quantitative
analysis, using newtonian gravity corrected for special
relativistic effects, which works surprisingly well, to
corroborate and illustrate/visualize these statements. This
article may be seen as a plea to start taking seriously the
observed inhomogeneous distribution and the nonlinearities of
nonperturbative general relativity, and their impact on the
dynamics and behavior of the cosmos.
\end{abstract}
Measurements since the late 1990s on type Ia supernovae (SN)
\cite{Riess},\cite{Perlmutter} surprisingly seemed to indicate
that the universe accelerates its expansion at the present epoch,
instead of the deceleration expected if gravity is universally
attractive. This finding has resulted in a ``neo-standard"
interpretation where the present universe is believed to be
dominated by ``dark energy" with negative pressure. However, when
taking the observed inhomogeneous structure of the universe into
account, and considering the real geodesic paths of observed SN
light, such a hypothesis might be superfluous.

i) If we assume that luminous matter (and by necessity gas, dust
and plasma for star formation) is a good ``tracer" of regions of
higher than average density, almost all photons from distant
objects that reach us on earth must have traversed regions with
little or no matter with which it can interact
electromagnetically. The light from distant objects ``zigzag"
through the maze defined by gravitationally bound objects
(galaxies, galaxy clusters, superclusters) making the real
(geodesic) path longer than the one calculated from standard,
perfectly isotropic and homogeneous, FRW-cosmology, the more so
the longer structure formation has progressed ($z \leq\sim 1$)
\cite{Ahlenius}. ii) The actual dynamics of the universe is
altered due to the self-induced inhomogeneities, again more
prominently so as structure formation progress. We will,
deliberately from a pedagogical standpoint, employ a heuristic
model using newtonian gravity, with special relativistic
corrections, both for the ease of visualization/interpretation and
the much simpler (linear) mathematics than full-blown (nonlinear)
general relativity, in which not even the two-body case is
analytically solvable. Even though newtonian gravity is not
completely mathematically consistent in an infinite universe, due
to its dependence on boundary conditions infinitely far away, we
can side-step this by treating the Big Bang as an "explosion" in a
pre-existing (newtonian) space. As can be seen in Fig. 2, this
special relativistically corrected newtonian model comes
surprisingly close to the general relativistic FRW-model,
especially for low $z$ as expected.

In general relativity, gravitationally bound systems have a
positive spacetime curvature. At the same time we know, from
observations of the cosmological microwave background radiation
(CMBR)\cite{Balbi},\cite{deBernardis}, that the global geometry of
the universe most probably is flat. This means that the curvature
between gravitationally bound systems (solar systems, galaxies,
galaxy clusters, etc) must be negative. This conclusion applies to
all globally flat universes with (semi-)localized gravitationally
bound systems.

For a truly exact description, we would need to know the
energy-momentum tensor ($T_{\mu \nu}$) at each point between us
and the distant SN, which is physically impossible. And even if we
had such perfect information, it would still be mathematically
inconceivable to solve the resulting Einstein equations,
\begin{equation}
R_{\mu \nu} - \frac{1}{2} g_{\mu \nu} R = \kappa \, T_{\mu \nu},
\end{equation}
to deduce the local curvature at each point in terms of the
Riemann curvature tensor, due to the complexity of the equations -
ten coupled, nonlinear PDEs. (The real lure of assuming perfect
homogeneity and isotropy is that Einstein's eqs. simplifies to two
coupled \textit{linear} ODEs, Friedmann's eqs., in which the total
dynamical behavior of such FRW-model universes is contained solely
in $a(t)$, the cosmic scale factor.)

One hope would be that an approximate model could be used to
determine an ``effective" curvature for bound systems and for the
space in-between. It would define a semi-local mean curvature
parameter, $k$, for the regions \textit{bound/between}. (This
parameter is related to the scalar curvature $R = R_{\mu}^{\mu}$,
averaged over the region, $\langle R \rangle$.) Even in such a
``coarse-grained" inhomogeneous model, the cosmic scale factor
$a(t)$ will be scale-dependent, as noted already by de
Vaucouleurs\footnote{``This leads one to view the Hubble parameter
as a stochastic variable, subject in the hierarchical scheme to
effects of local density fluctuations on all scales."} \cite{deV}.

In terms of $\Omega = \rho / \rho_{crit}$ (where $\rho_{crit}$ is
the density required for flatness):
\begin{equation}
\Omega_{global} = 1,
\end{equation}
\begin{equation}
\Omega_{bound} > 1,
\end{equation}
\begin{equation}
\Omega_{between} < 1.
\end{equation}
Or, stated in terms of the mass-energy density: $\rho_{global} =
\rho_{crit}$, $\rho_{bound} \gg \rho_{crit}$, $\rho_{between} \ll
\rho_{crit}$. However, one should keep in mind that the very
definition and usage of $\Omega$ assumes homogeneity and isotropy.

As matter preferentially clumps in well localized objects (stars,
etc), the majority of the photons that reach us travel mainly in
``under-dense" (negatively curved) $\Omega < 1$ space. By
observing \textit{light} we are thus automatically biased to
measure an ``apparent" curvature which is less than the actual
global curvature.\footnote{Neutrinos and gravitational waves
should show less bias as they can travel unhindered through huge
amounts of matter without interacting appreciably. This means that
both negative and positive curvature effects contribute, which
reduces the bias.} Light from a SN source will, in a universe
which exhibits gravitational clumping/structure formation, always
be switched towards a seemingly more negatively curved universe.
For a universe with zero global curvature, the SN light will thus
approach the curve for an \textit{open} universe, see Fig. 1.
\begin{figure}
\begin{center}
\scalebox{.5}{\includegraphics {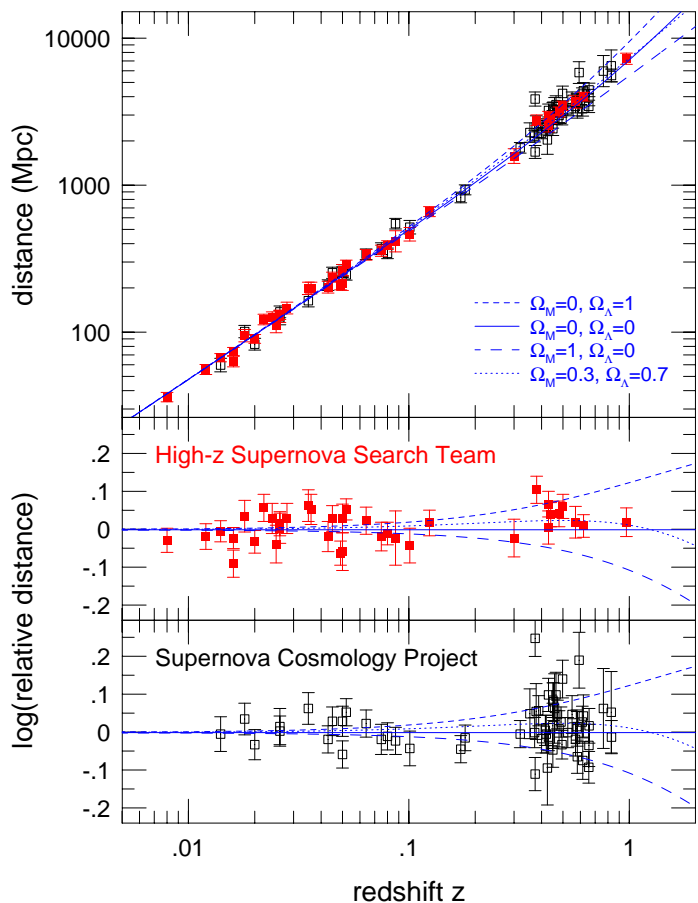}}
\end{center}
\caption{\small{Shown \cite{Sollerman} are the original data
points of the \textit{High-$z$ Supernova team} (filled
squares)\cite{Riess}, and the \textit{Supernova Cosmology Project}
(open squares)\cite{Perlmutter}. The dashed line is the
theoretical prediction for a homogeneous and isotropic (FRW)
universe which is flat and without cosmological constant
($\Omega_M = 1$, $\Omega_{\Lambda} = 0$). The solid line is the
corresponding prediction for an empty (open) universe ($\Omega_M =
0$, $\Omega_{\Lambda} = 0$). Also shown (short dashes) is the
theoretical prediction for a flat universe with solely a dark
energy component ($\Omega_M = 0$, $\Omega_{\Lambda} = 1$). The
dotted line is the solution currently favored for the SN Ia data
by both experimental groups ($\Omega_M = 0.3$, $\Omega_{\Lambda} =
0.7$). For $z>1$, where unfortunately also observational
measurements become increasingly difficult, it starts to deviate
towards the ($\Omega_M = 1$, $\Omega_{\Lambda} = 0$) line. The
neo-standard explanation for this is that $\Omega_{\Lambda}$ has
become dominant only fairly recently ($z \leq 1$). Observed SN
photons within a globally flat universe will always tend towards
the line for the open universe, due to inhomogeneous structure
formation (assuming $\Omega_{\Lambda} = 0$).}}
\end{figure}

Another compelling property is that this negative curvature effect
will automatically mimic a very small cosmological constant,
beginning to ``dominate" at an epoch when a significant amount of
structure has evolved. Before structure formation through
gravitational condensation becomes effective ($1100 \gg z
> 4$), all space will have roughly the same curvature ($k \simeq
0$). However, structure formation will produce bound systems with
increasing $\Omega_{bound}$, which means that $\Omega_{between}$
will be a decreasing function of time. Hence, the space in-between
bound systems will asymptotically approach $\Omega = 0$ with time
(under-density being diluted by expansion), simulating an
accelerated expansion.

After these general considerations, we now turn to the newtonian
model to qualitatively and semi-quantitatively, make our case
about ii), the dynamics. First, it is possible to deduce
analytical results from a maximally inhomogeneous and anisotropic
distribution - a two-body problem: \vspace{3mm}
\newline Nomenclature:\\
$d$, the distance from an object to the observer.\\
$v$, the observed velocity of an object in the radial direction.\\
$a$, the observed acceleration of an object in the radial direction.\\

We first define the "homogeneous acceleration", denoted $a_h$.
This is the acceleration that would be experienced by an object if
the distribution would have been homogeneous and isotropic on all
scales. Simply put, if a test particle is somewhere in a
homogeneous sphere, the only net effect of gravitation is the mass
within a smaller sphere with a radius equal to the distance from
the test mass to the center. The acceleration for this test mass
becomes
\begin{equation}
a_h = -\frac{G m}{d^2},
\end{equation}
where $m$ is the mass of the small sphere. This mass can be
expressed by the volume ratio times the total mass of the
distibution, $M$,
\begin{equation}
m = M\frac{d^3}{R^3},
\end{equation}
where $R$ is the radius of the total volume. The acceleration can
be expressed as
\begin{equation}
a_h = -\frac{G M d}{R^3}.
\end{equation}
If we know the total mass, then one way of deciding $R$ is to
check the mean-distance of the observed supernovas. If we assume
that we can see all supernovas in the observable universe, we can
then just take the mean-value of $d$ and multiply with four-thirds
to get the radius, as the geometric center lies at three-fourths
of the radius in a cone-fragment of a homogeneous sphere. The
equation for the acceleration becomes
\begin{equation}
a_h = -\frac{27 G M d}{64 \langle{d}\rangle^3},
\end{equation}
where $\langle{d}\rangle$ is the observed mean-value of the distance, $d$.

Another way is if we know the mean-density, $\rho$. Then the acceleration becomes
\begin{equation}
a_h = -\frac{4 \pi \rho G d}{3}.
\end{equation}

Let us now obtain a quantitative measure of inhomogeneity. If we
start with the acceleration, we can simply add and subtract the
``homogeneous acceleration", $a_h$,
\begin{equation}
a = a_h+\left(a-a_h\right),
\end{equation}
extract a factor $\frac{v^2}{d}$ from the bracket
\begin{equation}
a = a_h+\frac{v^2}{d}\left(\frac{a d}{v^2}-\frac{a_h d}{v^2}\right).
\end{equation}
Introducing the Hubble parameter, $H=\frac{v}{d}$, we get
\begin{equation}
a = a_h+d H^2\left(\frac{a d}{v^2}-\frac{a_h d}{v^2}\right).
\end{equation}
The terms inside the bracket are dimensionless. As can be seen,
this inhomogeneous result completely without dark energy yields
the same behavior as the standard (homogeneous) model with a
cosmological constant,
\begin{equation}
a = a_h+d H^2 \Omega_\Lambda.
\end{equation}
where instead of the bracket one has the $\Omega_\Lambda$ term
\cite{Peebles}. From now on the correction term for the
inhomogeneity will be denoted by
\begin{equation}
Q = \left(\frac{a d}{v^2}-\frac{a_h d}{v^2}\right).
\end{equation}

Let us consider an inhomogeneous Universe consisting of two large
bodies and an observer, whose mass is negligible, situated between
them. The two bodies have equal mass and therefore half the mass
of the Universe.

As this is a two-body problem the acceleration in the radial
direction is
\begin{equation}
a = -\frac{G M}{8 d^2}.
\end{equation}
The gravitational potential, $V$, is defined according to
\begin{equation}
F = -\bigtriangledown_{2d} V,
\end{equation}
where $F$ is the force acting on the particles in the radial
direction,
\begin{equation}
F = -\frac{G M^2}{16 d^2}.
\end{equation}
Solving for $V$ gives
\begin{equation}
V = -\frac{G M^2}{8 d}.
\end{equation}
If the particles have escape velocity (corresponding to flat space-time), the virial theorem states that
\begin{equation}
2K + V = 0,
\end{equation}
where $K$ is the kinetic energy of one of the particles, $K = -\frac{M v^2}{4}$. Solving for $v^2$ gives
\begin{equation}
v^2 = \frac{G M}{4 d}.
\end{equation}
The total mass of the Universe is $M$ so,
\begin{equation}
a_h = -\frac{27 G M d}{64 \langle{d}\rangle^3}.
\end{equation}
In this case $\langle{d}\rangle = d$ because both particles are at
the same distance from the observer. So the "homogeneous
acceleration" becomes
\begin{equation}
a_h = -\frac{27 G M}{64 d^2},
\end{equation}
which gives the value,
\begin{equation}
Q = \left(-\frac{1}{2}+\frac{27}{16}\right) = 1.1875.
\end{equation}
Thus the real acceleration will be
\begin{equation}
a = a_h + d H^2 1.1875.
\end{equation}

The motivation for this somewhat trivial demonstration is to show
that if we observe an inhomogeneous Universe and still ``pretend"
it is homogeneous, it will yield a correction term for the
inhomogeneity that behaves in the same way as a cosmological
constant. This points to a possibility to remove the need for a
mysterious dark energy which today has no fundamental, microscopic
explanation or justification whatsoever. \vspace{3mm}
\newline An N-body simulation with newtonian gravitation that
takes into account the effects of special relativity has also been
done. The particles have a small gas-like extension so if they get
really close to each other the gravitational potential will not be
infinite. Instead of having the ordinary potential going like
$-\frac{1}{d}$ (where $d$ now is the distance between two
particles), it will go like $d^2$ if the particles are closer than
their radius, $r$, given at the start of the simulation. If the
particles are separated by a distance larger than $r$ there will
be a potential like
\begin{equation}
V = -G\frac{m^2}{d},
\end{equation}
where $m$ is the mass of each particle, while for $d < r$ it is
\begin{equation}
V = -G\frac{3m^2}{2r}+G\frac{m^2d^2}{r^3}.
\end{equation}
The particles will therefore not collide but simply pass through
each other without having an infinite acceleration, just as
expected for the ``test-particles" of cosmology; galaxies. They
are distributed randomly in a sphere so that the large-scale
density is almost uniform. The particles begin with escape
velocity as initial condition, as this is equivalent to a flat
Universe in the Einstein setting. Then the acceleration is
calculated and the next distance and velocity for all particles
are iterated using a finite difference method. Plotting the
distance from the center of the sphere against red-shift, and
comparing to a few general relativistic ``standard" cosmological
models, show some interesting results, Fig. 2. All models give
similar behavior up to $z \sim 1$ and the particles in the flat
but inhomogeneous model approach the homogeneous \textit{open}
model for higher $z$, in line with our previous analytical result.
\begin{figure}
\begin{center}
\scalebox{.5}{\includegraphics {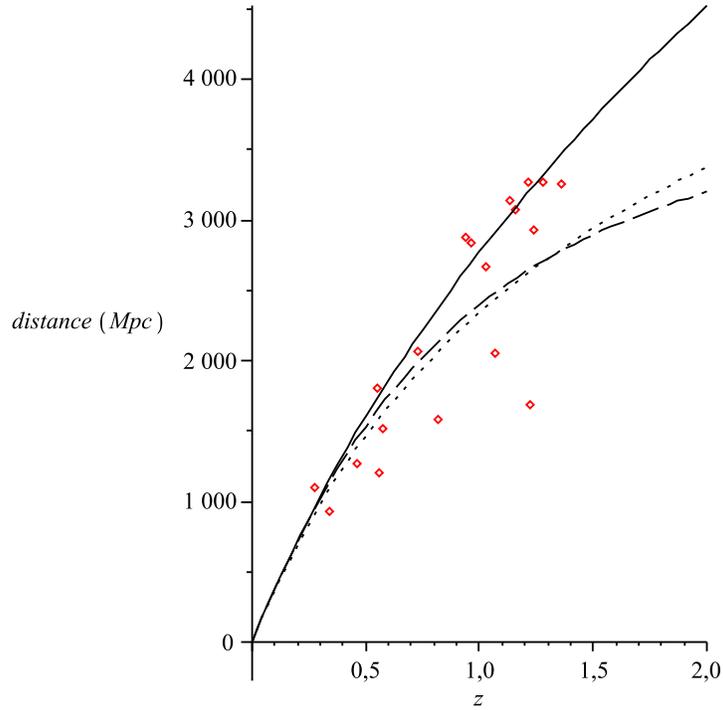}}
\end{center}
\caption{\small{The prediction of three different homogeneous
models are shown together with our simulated inhomogeneous
``particles". The solid line is an open FRW-model with $\Omega_M =
0.2$, $\Omega_\Lambda = 0$. The dotted line is a flat FRW-model
with $\Omega_M = 1$, $\Omega_\Lambda = 0$. The dashed line is the
(flat) newtonian prediction with corrections for special
relativity. An inhomogeneous matter distribution may thus be
interpreted wrongly as to suggest that $\Omega_M$ is lower than it
actually is, \textit{e.g.} $\Omega_M = 0.2$ if interpreted through
a homogeneous and isotropic ``standard model" cosmology. The
simulated particles correspond to $\Omega = \Omega_M = 1$. The
assumption of homogeneity and isotropy mislead the physical
interpretation if the real distribution is inhomogeneous, as is
the case for our Universe.}}
\end{figure}
\vspace{3mm} \newline We end with some related comments:
\begin{itemize}

\item SN Ia data probe regions with $z \leq 1.7$.
 Homogeneity and isotropy is valid only on scales significantly
 larger (orders of magnitude) than the cosmological ``voids" and ``filaments" \cite{Einasto1,Einasto2},
 \textit{i.e.}, at distances $\gg$ 120 Mpc (corresponding roughly to $z \sim 0.03$).
 Neither can one \textit{a priori} rule out clumping on even
 grander scales. Hence, the ``cosmological principle" of
 homogeneity and isotropy, which the FRW-solution crucially depends upon,
 does not apply exactly. Instead full consideration of the
 inhomogeneities should be taken, at least up to the distance scale where homogeneity and isotropy
 may be considered a valid approximation.

\item The CMBR almost certainly probes the overall
geometry/curvature of the universe ($z \sim 1100$), as little
gravitational structure could form/grow before photon decoupling.
The statistical weight of the low $z$-range where appreciable
structure has formed is negligible compared to the higher
$z$-range which thus dominates the integrated effect for the CMBR.
For very high redshift the photons accordingly should behave ``as
expected" in a flat universe. Also, the CMBR is ``everywhere"
while SN photons travel from a pointlike source to us along a
sharp geodesic ``ray". This means that, due to the inhomogeneity
and anisotropy at small to medium scales, constraints from SN and
CMBR may not ``carry over" trivially between one another.

\item In a FRW-universe with a cosmological constant it is just a
strange and completely unexplained ``cosmic coincidence" that
$\Omega_{M} \sim \Omega_{\Lambda}$ \textit{now} ($\Omega_{M} \gg
\Omega_{\Lambda}$ earlier and $\Omega_{M} \ll \Omega_{\Lambda}$
later). However, in our scenario it is an automatic bonus, as an
appreciable amount of structure must form before intelligent life
can evolve to observe it. It is thus natural that we live in an
epoch when the apparent ``acceleration'' (really due to
inhomogeneity) becomes observable.

\end{itemize}

In conclusion, we have noted that by regarding the real
inhomogeneous matter distribution arising from time-dependent
gravitational structure formation, it might be possible to avoid
the conclusion that the expansion of the universe accelerates, as
normally drawn from high-$z$ SN Ia data. This would alleviate the
need to postulate that the present universe at large is dominated
by an exotic ``dark energy" with a mysterious negative pressure.

This can also be seen as a plea to start taking seriously the
nonlinear character of general relativity, one of the
intrinsically most highly nonlinear theories in existence, where
even very ``small" perturbations could grow exponentially. This,
and other phenomena, are known to occur in other (e.g. chaotic)
nonlinear systems, whereas overly simplified linearized models, or
even the whole perturbative series with infinitely many terms,
suppress or even exclude such behavior. Coupled with the vast
distances and timescales relevant in cosmology this could make all
the difference.

For example, a FRW-model with newtonian perturbations as used in
simulations of large-scale structure formation may be overly
simplified to capture the true dynamics. We believe that the
nonlinear aspects of gravity have been gravely underestimated in
cosmology, including very important but unsolved turbulent
processes.

Despite high-precision observations advancing cosmology towards
becoming a normal ``exact science" in the last couple of decades
the present physical understanding of the universe as a whole is
probably still very crude. We predict that theoretical cosmology
fifty years from now will bear little resemblance to the
FRW-models almost universally used today.

\end{document}